\documentclass[aps,pre,preprint,superscriptaddress]{revtex4-1}

\usepackage{graphicx,amssymb,verbatim,amsmath,siunitx,color,float, soul}
\usepackage{natbib,url, enumerate}
\usepackage[colorlinks,allcolors = blue,bookmarksopen,bookmarksnumbered]{hyperref}
\usepackage{tabularx}
\usepackage[table]{xcolor}
\usepackage{multirow}

\usepackage{hyperref}

\begin{document}

\title{Coarsening Kinetics in Active Model B+: Macroscale and Microscale Phase Separation}

\author{Pradeep Kumar Yadav}
\affiliation{School of Physical Sciences, Jawaharlal Nehru University, New Delhi -- 110067, India.}
\author{Shradha Mishra}
\affiliation{Department of Physics, Indian Institute of Technology, BHU  -- 221005, India.}
\author{Sanjay Puri}
\affiliation{School of Physical Sciences, Jawaharlal Nehru University, New Delhi -- 110067, India.}

\begin{abstract}
We perform a comprehensive numerical investigation of the coarsening kinetics of active Brownian particles modeled by the {\it Active Model B+} (AMB+). This model was introduced by Tjhung et al. [Phys. Rev. X {\bf 8}, 031080 (2018)] and is a generalization of Model B for a conserved order parameter, with two additional activity terms. These terms correspond to rotation-free current (of strength $\lambda$) and rotational current (of strength $\xi$). There is a range of $(\lambda, \xi)$-values for which the system undergoes macroscale phase separation (MPS). In this case, the order parameter current develops a structured pattern consisting of alternating clockwise and anticlockwise loops along domain walls. These pairs are separated by nodal points where the current magnitude is very small. The mass transfer is driven by the overlap of loops on neighboring droplets. Thus, the mass has to undergo circular trajectories while transferring from smaller droplets to larger droplets. This slows down domain growth in comparison to Model B, where the current is transported by the shortest route between droplets. Thus, there is slower coarsening in AMB+, with an asymptotic growth exponent of $1/4$. For another range of $(\lambda, \xi)$-values, the system undergoes microscale phase separation ($\mu $PS), i.e., it reaches a steady-state morphology characterized by a crystal of monodisperse droplets, whose size depends on the parameters. We present detailed results for the kinetics of MPS and $\mu$PS in AMB+ with a critical composition, where the system undergoes spinodal decomposition.
\end{abstract}

\maketitle

\section{Introduction}
\label{s1}

Active matter is an assembly of motile particles that dissipate energy at the microscopic level to self-propel~\cite{Marchetti_2013, Cates2015}, e.g., molecular motors, actin filaments, microtubules, fish schools, bird flocks, and autophoretic colloids~\cite{calovi2014swarming, Buttinoni2013, ballerini2008}. The most natural forms of active matter are biological systems~\cite{shaebani2020computational}. Active matter is intrinsically nonequilibrium and exhibits many intriguing phenomena, such as collective behavior, motility-induced phase separation (MIPS) and giant number fluctuations~\cite{narayan2007long,Marchetti_2013,shaebani2020computational}.For example, the rich phenomenology observed in dense bacterial suspensions arises from a combination of self-propulsion and alignment interactions due to their rod-like shapes~\cite{sokolov2007concentration}. Vicsek and others focused on motile particles with orientational order, including active nematics, self-propelled rods, and active Ising spins \cite{vicsek1995novel, solon2013revisiting, ramaswamy2010mechanics}. Their studies reveal that the interplay between the alignment interaction and self-propulsion yields a variety of phases in these systems \cite{Marchetti_2013,vicsek2012collective}.

Another intriguing class of systems consists of active colloids that lack orientational order, such as motile colloids with spherical shapes (e.g. synthetic Janus particles). Alignment interactions in these systems are usually negligible and particles can be considered  geometrically isotropic, in contrast to elongated or rod-like active matter. These particles with purely repulsive interactions undergo liquid-gas phase separation, which is not possible for passive colloids without attractive forces. This process is known as MIPS and is the result of a nonequilibrium mechanism due to intrinsic motility \cite{Stenhammar2013,cates_tjhung_2018,caporusso2020motility,mandal2019motility,caporusso2020motility,shi2020self,martin2021characterization}. For example, bacteria with quorum sensing and active Janus particles with density-dependent self-propulsion speed exhibit MIPS~\cite{worlitzer2021motility,sokolov2007concentration}. The underlying mechanism is due to the lack of microscopic time-reversal symmetry (TRS) in these systems. Consequently, the steady state of such systems does not obey the principle of detailed balance.

Numerous studies using different microscopic models based on {\it active Brownian particles} (ABP) or {\it run-and-tumble particles} (RTPs) have elucidated the dynamical and steady-state properties of MIPS~\cite{Jailleur2008, cates2013active, Stenhammar2013,Fily2012, Jailleur2008, Cates2015, Stenhammar2013, Buttinoni2013}. In addition, continuum field theories have also been developed to explore the complex phases emerging from phase separation in active systems~\cite{shaebani2020computational,Cates2015}.
These theories provide a framework for relating microscopic parameters to macroscopic behaviors, thus facilitating experimental control over the novel phases of active matter~\cite{shaebani2020computational, Stenhammar2013, Cates2015}. The simplest continuum models describing MIPS in an active colloid are generalizations of the well-known {\it Cahn-Hilliard} (CH) equation \cite{ch58} or {\it Model B} (MB) \cite{modelb}, which describes phase separation in passive systems. These
are called {\it Active Model B} (AMB)~\cite{Wittkowski2014} and AMB+~\cite{Tjhung2018}. We will discuss the mathematical formulations of AMB and AMB+ shortly.

In this paper, we focus on coarsening kinetics in AMB+. This problem is well studied in the context of MB and has received a great deal of attention in the literature~\cite{pw09}. In the classical problem, a homogeneous (disordered) binary mixture is rendered thermodynamically unstable by a rapid quench below the critical temperature. The unstable system evolves through the emergence and growth of domains enriched in the different components of the mixture. This domain growth is characterized by a single time-dependent length scale $L(t)$ that follows a power-law behavior, i.e., $L(t) \sim t^{1/z}$, where $z$ is the dynamic growth exponent. For MB, the coarsening is driven by diffusive transport, which yields the Lifshitz-Slyozov (LS) growth law with $z=3$ \cite{pw09}. The pattern dynamics of the far-from-equilibrium system is characterized by \textit{dynamical scaling} of the correlation function and the structure factor. In physical terms, this corresponds to statistical self-similarity of the morphology with time: the only feature that changes is the domain size $L(t)$.

In an earlier work~\cite{sss21}, we conducted a study of segregation kinetics in AMB \cite{Wittkowski2014}. AMB is an active version of MB and includes a term (of strength $\lambda$) that is not derivable from a free energy functional. Our numerical results for AMB showed two important features: (a) The growth law shows a crossover from the LS law to a novel growth law (consistent with $L(t)\sim t^{1/4}$) at late times. The crossover time has an inverse dependence on the activity strength $\lambda$. (b) The asymptotic correlation function exhibits dynamical scaling for each $\lambda$, but the scaling function is not superuniversal and depends on $\lambda$. 

The present paper extends our previous work to the case of AMB+, proposed by Tjhung et al.~\cite{Tjhung2018}. This is a further generalization of AMB to a case with two distinct terms which model activity. These additional terms cannot be derived from a free energy as they are intrinsically nonequilibrium terms. As in our earlier study, we focus on the domain growth kinetics in AMB+. We are particularly interested in the dynamical interplay of the two activity terms.
 
This paper is organized as follows. In Sec.~\ref{s2}, we introduce AMB and AMB+, and present details of our simulation study. In Sec.~\ref{s3}, we present numerical results for domain growth in AMB+. Finally, in Sec.~\ref{s4}, we conclude this paper with a summary and discussion of our results.

\section{Model and Simulation Details}
\label{s2}

Coarsening kinetics in passive systems (e.g. binary mixtures without activity) is described by a dynamical field theory with a conserved order parameter field, $\psi({\bf r}, t)$, where ${\bf r}$ denotes space. Here, $\psi$ represents the local concentration difference between the two species. These systems obey detailed balance and preserve TRS. For passive systems with diffusive transport, the evolution of $\psi$ obeys the continuity equation or MB~\cite{modelb}:
\begin{eqnarray}
    \frac{\partial \psi({\bf r},t) }{\partial t}  &=& -{ \nabla}\cdot {\bf J} ({\bf r},t), \label{continuity} \\
     {\bf J}({\bf r},t) &=& -M\nabla \mu ({\bf r},t) = -M\nabla \left( \frac{\delta{\mathcal{F}[\psi]}}{\delta \psi}\right). 
     \label{current}
\end{eqnarray}
Here, ${\bf J}({\bf r},t)$ is the current density, $\mu({\bf r},t)$ is the local chemical potential, and $M$ is the mobility which is generally taken to be independent of $\psi$. In general, the current also contains a random noise term. In Eq.~(\ref{current}), we consider the deterministic case with zero noise. For MB, it is well known that thermal noise is asymptotically irrelevant to the domain growth process \cite{po88,ab89}. The quantity ${\mu}$ can be derived from the relevant free energy functional $\mathcal{F}[\psi]$, which is generally modeled by the Ginzburg-Landau $\psi^4$-form: 
\begin{eqnarray}
  \mathcal{F}[\psi({\bf r},t)] &=& \int{d {\bf r}\left\{\frac{a(T-T_c)}{2}\psi^{2} + \frac{b}{4}\psi^{4} + \frac{K}{2}\left|{\nabla}\psi\right|^{2}\right\}}.
\label{landauen} 
\end{eqnarray}

In Eq.~(\ref{landauen}), $a$, $b$, and $K$ are positive constants, and $T$ is the temperature. The critical temperature of phase separation is denoted as $T_c$. The first two terms in the RHS represent the bulk free energy density. The third term represents the energy cost (surface tension) associated with variations in the order parameter. For $T<T_c$, MB governs the evolution of $\psi$ to the free energy minima
\begin{equation}
\psi_0 = \pm \left[\frac{a}{b} (T_c-T)\right]^{1/2} .
\end{equation}
For a general form of the free energy, the equilibrium phases are obtained from the well-known Maxwell common tangent construction, which ensures the same chemical potential and pressure between the coexisting phases.

Active matter violates detailed balance: thus, TRS, which is fundamental to MB, does not hold for active particles. Thus, an extension of MB with terms that explicitly break TRS is required to describe phase separation in active systems. The first nonlinear terms that possess this property arise as a correction to the dynamical equation of $\psi$ at $\mathcal{O}(\nabla^4 \psi^2)$. (Note that any term that can be absorbed by redefining the free energy functional $\mathcal{F}$ cannot break TRS.) Therefore, the minimal field theory for AMB+ includes terms up to the order $\mathcal{O}(\nabla^4 \psi^2)$~\cite{Tjhung2018, nardini2017entropy}. These terms can be of two types: (i) $\nabla \cdot \nabla (|\nabla \psi|)^2$, which is associated with zero curl current, and (ii) $\nabla \cdot (\nabla^2 \psi \nabla \psi)$, which is associated with nonzero curl current.

The governing equation of AMB+ for the phase separation of active particles is given by~\cite{Tjhung2018}
\begin{equation}
    \frac{\partial \psi({\bf r},t) }{\partial t}  = -\nabla \cdot {\mathbf J} ({\bf r},t) = -{ \nabla}\cdot \left[{-\nabla \left(\mu_{\rm eq} + \lambda |\nabla \psi|^2 \right)+ \xi  {{\nabla} ^2}\psi{\nabla}\psi}\right], \label{ambcontinuity}
\end{equation}
where $\lambda$ and $\xi$ are activity parameters. The term $\lambda|\nabla \psi|^2 = \mu_{\rm neq} ({\bf r},t)$ defines the local nonequilibrium chemical potential, which is added to the equilibrium chemical potential $\mu_{\rm eq} ({\bf r},t) = a(T-T_c)\psi + b\psi^3-K\nabla^2\psi$ \cite{Wittkowski2014}. A similar term arises in the Hunter-Saxton equation for nematic liquid crystals~\cite{HSE} and the Kardar-Parisi-Zhang equation for nonlinear interfacial diffusion~\cite{KPZ}.  The case with $\lambda \ne 0$ and $\xi=0$ is termed AMB by Wittkowski et al., and was previously studied by us~\cite{sss21}. AMB does not take into account circulating currents, which may be important for active colloids~\cite{Stenhammar2013, Diptabrata2023}, and also biological active matter such as dense active nematics \cite{prl2023}, bacterial baths \cite{pnas2009}. The last term (with strength $\xi$) in Eq.~\eqref{ambcontinuity} represents the circulating current.

For MB, we can use the natural scale of the order parameter, space, and time to rescale it to a parameter-free form~\cite{pw09}. For $T<T_c$, the corresponding dimensionless form of AMB+ is
\begin{eqnarray}
    \frac{\partial \psi({\bf r},t) }{\partial t}  &=& -\nabla \cdot {\mathbf J} ({\bf r},t) \\
    &=& -{\nabla} \cdot \left[{-\nabla}\left( -\psi + \psi^{3} -\nabla^{2}\psi +  \lambda^{\prime}{\left|{\nabla} \psi\right|}^{2} \right) + \xi^{\prime}{{\nabla} ^2\psi{\nabla}\psi} \right].
    \label{MBplus}
\end{eqnarray}
In Eq.~\eqref{MBplus}, the parameters $\lambda^{\prime}$ and $\xi^{\prime}$ denote the rescaled counterparts of $\lambda$ and $\xi$. For $\lambda' = \xi' = 0$, we obtain the dimensionless MB with the fixed-point solutions $\psi_0 = \pm 1$. In our subsequent discussion, we will drop the primes for convenience, that is, $\lambda^{\prime} \to \lambda, \xi^{\prime} \to \xi$.

The phase diagram of Eq.~(\ref{MBplus}) in the $(\lambda, \xi)$-plane is shown in Fig.~4 of Ref.~\cite{Tjhung2018}. There is a region of ($\lambda,\xi$)-values where the system settles into steady-state droplet configurations with monodisperse sizes. This region corresponds to {\it reverse Ostwald ripening}, where smaller droplets grow at the expense of larger ones. Here, the density outside the larger droplets exceeds that around the smaller droplets, causing mass flux from large to small. In contrast, there is another region of ($\lambda,\xi)$-values where the system undergoes {\it forward Ostwald ripening}, with smaller droplets growing at the expense of larger ones, eventually leading to bulk phase separation.  

It is useful to find the coexisting phase-separated solutions of Eq.~\eqref{MBplus}. This is done by examining the static 1-$d$ kink solutions $\psi_{s}(z)$, which obey
\begin{equation}
\frac{d}{dz} \left[ -\psi_s + \psi_s^{3} - \frac{d^{2}\psi_s}{dz^{2}} + \lambda \left( \frac{d\psi_s}{dz} \right)^{2} \right] - \xi~\frac{d^{2}\psi_s}{dz^{2}}~\frac{d\psi_s}{dz} = 0.
\end{equation}
Notice that the $\lambda$- and $\xi$-terms are the same in $d=1$, as expected. Thus, the static kink obeys~\cite{Wittkowski2014,sss21}:
\begin{equation}
  -\psi_s + \psi_s^{3} - \frac{d^{2}\psi_s}{dz^{2}} + \left(\lambda-\frac{\xi}{2}\right) \left( \frac{d\psi_s}{dz} \right)^{2} = \mu_s,
\end{equation}
where $\mu_s$ (the static chemical potential) is nonzero for $\alpha=\lambda-\xi/2 \neq 0$. For small $\alpha$, a perturbative calculation yields~\cite{sss21}
\begin{eqnarray}
    \mu_{s}(\alpha) &=& \frac{4}{15}\alpha + \mathcal{O}(\alpha^2).
    \label{mus}
\end{eqnarray}
In addition, the saturation values of the order parameter (as $z \rightarrow \pm \infty$) obey the cubic equation:
\begin{equation}
-\psi_s + \psi_s^3 = \mu_s(\alpha) . 
\end{equation}
It is straightforward (though messy) to solve this cubic equation. The corresponding perturbative result is \cite{Wittkowski2014}
\begin{eqnarray}
    \psi_1 &=&1 + \frac{\mu_s}{2}, \nonumber\\
        \psi_2 &=&-1 + \frac{\mu_s}{2}.
        \label{psisat}
\end{eqnarray}
Clearly, for $\mu_s>0~(\alpha > 0)$, the $\psi>0$ phase ($\psi_1$) is more enriched than the $\psi<0$ phase ($\psi_2$). The opposite happens for $\alpha < 0$. The asymmetry in the two phases (for $\lambda \neq \xi/2$) is a novel feature of AMB+ and has interesting consequences for pattern dynamics.

We numerically solve Eq.~\eqref{MBplus} on a two-dimensional $N^2$ lattice ($N=512$) with periodic boundary conditions. The continuum field $\psi$ is discretized at the nodes $(x_i, y_i)$. Using the Euler discretization scheme, we solve the partial differential equation with a mesh size of $\Delta x = \Delta y = 0.5$ and a time step of $\Delta t = 0.01$, thus ensuring stability for the numerical scheme. We use the central-difference method for the space derivatives and the forward-difference method for the time derivative in Eq.~\eqref{MBplus}.

The initial condition consists of uniformly distributed small-amplitude fluctuations about $0$: $\psi({\bf r},0) = 0 + \delta\psi({\bf r},0)$, where $\delta \psi \in [-0.05, 0.05]$. This mimics the disordered state with critical composition prior to the ``quench''. The system is spontaneously unstable and undergoes phase separation via spinodal decomposition. The conservative nature of AMB+ ensures that $\int \psi({\bf r},t)~d{\bf r} = 0$ for all $t$. All other cases with a nonzero average of $\psi$ are known as {\it off-critical mixtures} and correspond to asymmetric binary mixtures.

We examine the effects of $\xi$ and $\lambda$ on the coarsening kinetics of active particles. Eq.~\eqref{MBplus} is invariant under the transformation $(\psi, \lambda, \xi) \to  (-\psi, -\lambda, -\xi)$. Therefore, the critical mixture yields the same results for the activity parameters $(\lambda, \xi)$ and $(-\lambda, -\xi)$. All statistical results presented here are averaged over 50 independent runs, except where otherwise mentioned.

\section{Numerical Results}
\label{s3}

\subsection{Macroscale Phase Separation (MPS)}
\label{3a}

\begin{figure}
\centering
\includegraphics[width=1.0\linewidth]{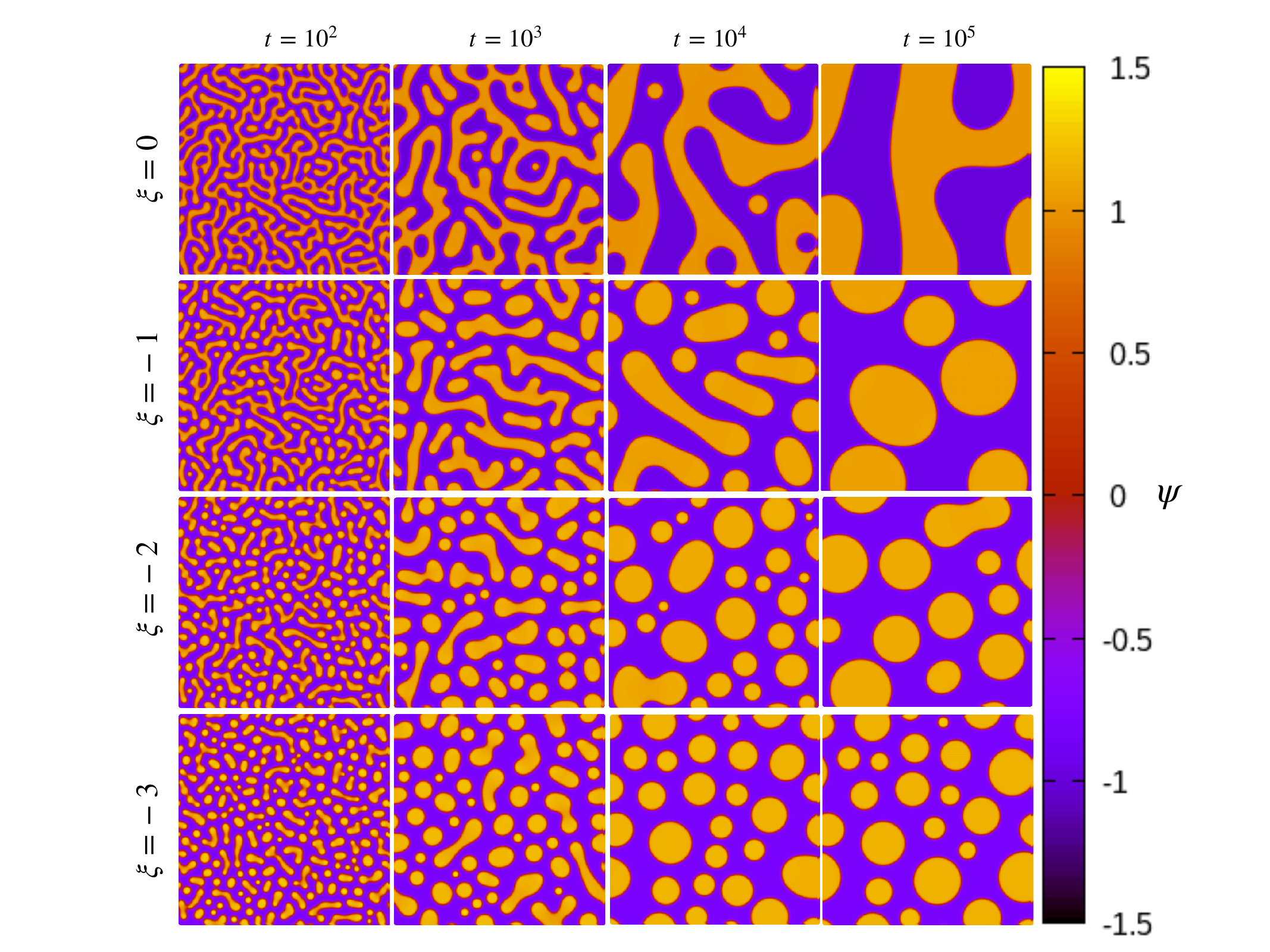} 
\caption{Evolution snapshots of AMB+ at $t = 10^2, 10^3, 10^4$, and $10^5$. The parameter values are $\lambda = 0$ and $\xi = 0, -1, -2$, and $-3$. The simulations are performed on a system of size $L^2 = 512^2$, with an average order parameter $\psi_0 = 0$, corresponding to a critical composition. The color bar represents the range of $\psi$-values.}
\label{f1}
\end{figure}

First, we study the evolving domain morphologies for $\lambda =0$ and different values of $\xi$. (We have already studied the case with $\xi=0$ and $\lambda$ varying in \cite{sss21}.)  The evolution snapshots of the local order parameter $\psi$ for $\lambda = 0$ and $\xi = 0, -1, -2, -3$ are shown in Fig.~\ref{f1} at different times $t =10^2$, $10^3$, $10^4$, $10^5$. The color bar shows the magnitude of $\psi$. These parameter values lie in the region labeled {\it forward Ostwald ripening} in the phase diagram of \cite{Tjhung2018}, referring to the conventional growth process where larger droplets grow at the cost of smaller ones. In contrast to MB, there is an asymmetry between the saturation values of $\psi$ for the two regions when $\xi<0$. We obtain a bicontinuous morphology for $\xi =0$, which is characteristic of MB (top row of Fig.~\ref{f1}). For $\xi < 0$, $\alpha = \lambda - \xi/2 > 0$ and $\psi_1$ is more enriched than $\psi_2$ -- see Eq.~(\ref{mus}). Thus, we see isolated droplets with $\psi > 0$ in a background of $\psi < 0$ at late times, even though we have a critical composition. However, the morphology at early times appears bicontinuous, as the domains are not saturated to their steady-state values. The evolution morphologies for $\xi>0$ are analogous to those of Fig.~\ref{f1}, except that the droplets correspond to the phase with $\psi < 0$.

To quantify the domain morphology, we measure the equal-time correlation function, defined as follows:
 \begin{eqnarray}
C({\bf r}, t) &= \langle \psi({\mathbf{R}},t) \psi({\mathbf{R}+\mathbf{r}},t) \rangle - \langle \psi({\mathbf{R}},t) \rangle \langle \psi({\mathbf{R}+\mathbf{r}},t) \rangle,
\label{crt}
\end{eqnarray}
where $\langle \ldots \rangle$ denotes an average over independent initial conditions. As our system is isotropic, $C({\bf r}, t)$ can be spherically averaged to obtain $C(r,t)$. This quantity exhibits dynamical scaling if domain growth is characterized by a unique length scale ${L(t)}$ \cite{pw09}:
\begin{eqnarray}
 C({r},t) &=& f\left(\frac{{ r}}{L}\right)= f(x),
\end{eqnarray}
where $f(x)$ is a scaling function. We define ${L(t)}$ as the distance at which ${C(r,t)}$ first decays to zero. A crucial quantity to understand the asymptotic growth law is the {\it effective dynamic exponent}, defined as
\begin{eqnarray}
{z_{\mathrm{eff}}} &=& \left[\frac{d (\ln L)}{d (\ln t)} \right]^{-1}.
\label{zeffective}
\end{eqnarray}

\begin{figure}
\centering
\includegraphics[width=0.85\linewidth]{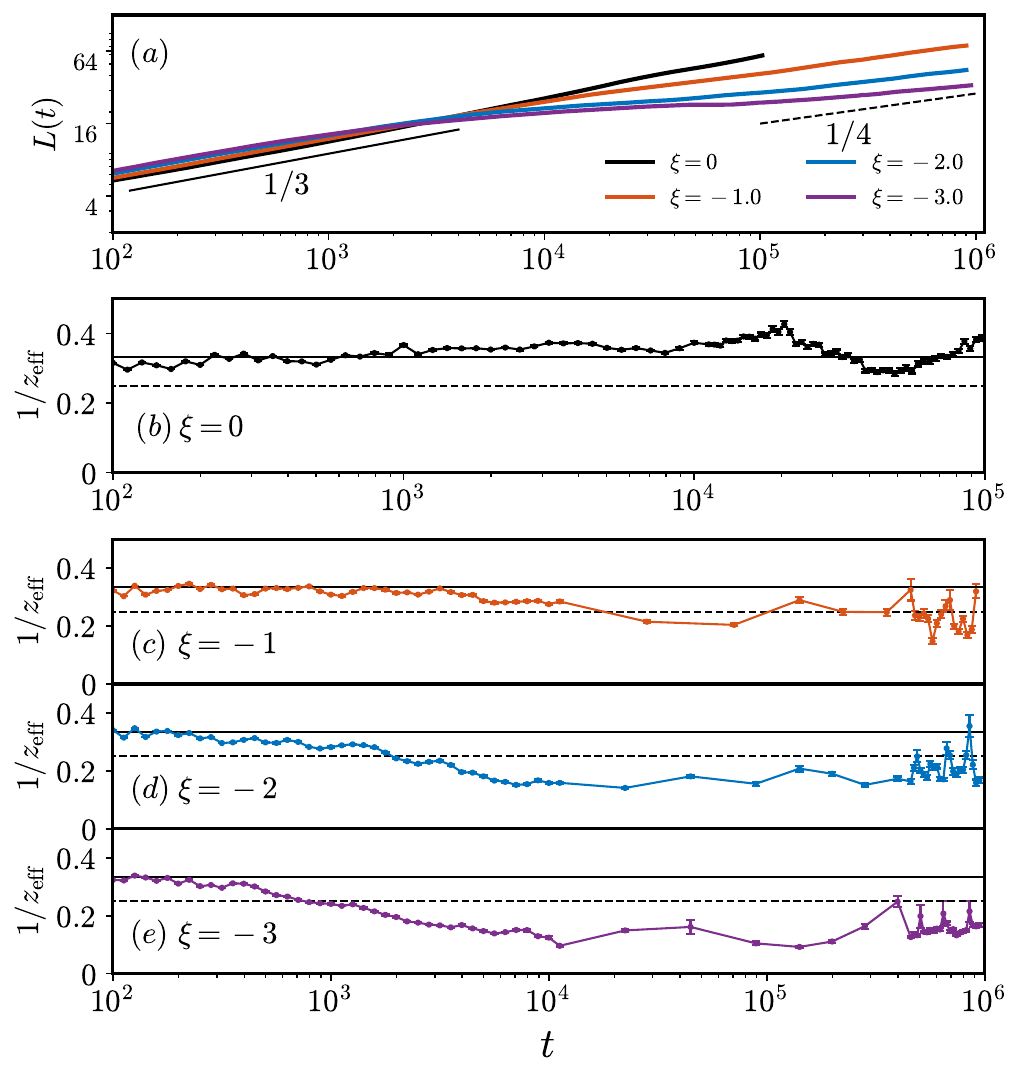} 
\caption{(a) Log-log plot of domain size $L(t)$ vs. time $t$ for $\lambda = 0$ and $\xi = 0,-1,-2,-3$, as labeled. The solid black line has a slope of $1/3$, and the dashed line has a slope of $1/4$. (b)–(e) Time-dependence of the inverse dynamic exponent $1/z_{\rm eff}$ for $\xi= 0,-1,-2,-3$.
The horizontal solid and dashed lines denote $1/z_{\rm eff} = 1/3$ and $1/4$, respectively. The error bars are computed from averages over the independent runs. At early times, the error bars are smaller than the symbol sizes. The time of the first crossing of the $1/z_{\rm eff} = 1/4$ line is designated as the crossover time $t_c$.}
\label{f2}
\end{figure}

In Fig.~\ref{f2}(a), we plot $L(t)$ vs. $t$ on a log-log scale for the evolution shown in Fig.~\ref{f1}. For $\xi = 0$, we observe the well-established LS power law, where ${L(t)} \sim t^{1/3}$. For $\xi < 0$, we observe an initial growth with the $1/3$ power law followed by a smooth transition to a slower growth law, which is consistent with $L(t) \sim t^{1/4}$ in the asymptotic regime. This is analogous to the behavior we have seen for AMB [Eq.~\eqref{MBplus} with $\xi=0$] in our earlier study~\cite{sss21}.

\begin{figure}
\centering
\includegraphics[width=0.8\linewidth]{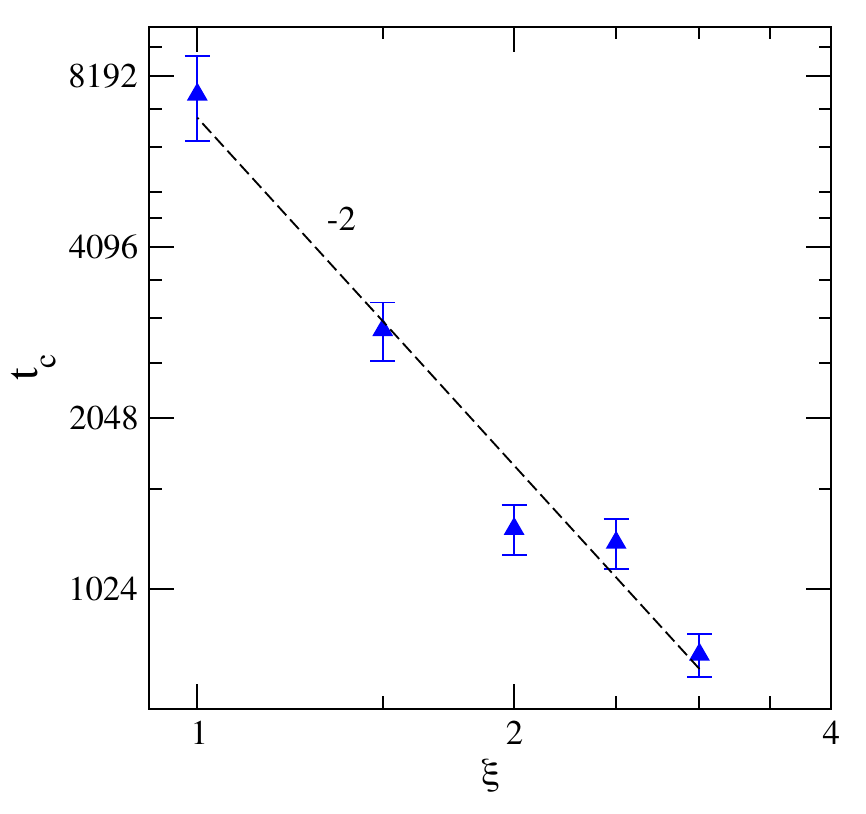} 
\caption{Log-log plot of the crossover time $t_c$ as a function of $\xi$. The dashed black line has a slope of $-2$.}
\label{f3}
\end{figure}

To quantify the asymptotic growth law, we calculate ${z_{\mathrm{eff}}}$ as a function of $t$. In Figs.~\ref{f2}(b)-\ref{f2}(e), we plot $1/z_{\mathrm{eff}}$ vs. $t$. For $\xi=0$, $z_{\mathrm{eff}}$ shows a flat trend with fluctuations around the expected value $1/z_{\mathrm{eff}}=1/3$. We should remark that the data for $z_{\rm eff}$ is usually noisy due to the numerical approximation of the derivative in Eq.~\eqref{zeffective}. For $\xi < 0$, at early times, all plots are flat around $1/z_{\mathrm{eff}}=1/3$, followed by a dip to values smaller than $1/4$ for $t>t_c (\xi)$.  At late times, all plots tend towards the exponent $1/4$. A couple of remarks are in order here. First, the crossover time $t_c$ has an inverse dependence on $\xi$. We define $t_c$ as the time where $1/z_{\rm eff}$ first falls to $1/4$ in Fig.~\ref{f2}(c)-(e), and plot $t_c$ vs. $\xi$ on a log-log scale in Fig.~\ref{f3}. The data show an approximate power law behavior with $t_c\sim \xi^{-2}$, although the range of $\xi$ is limited. Our second remark is that AMB+ shows $t^{1/3}$-growth in the window where the morphologies are bicontinuous in Fig.~\ref{f1}, whereas the system shows $t^{1/4}$-growth in the asymptotic regime where the morphology is droplet-like.  A similar droplet morphology arises for MB with off-critical composition $\psi_0 \neq 0$. However, in that case, the growth of the droplets continues to obey the LS law. 

\begin{figure}
\centering
\includegraphics[width=0.95\linewidth]{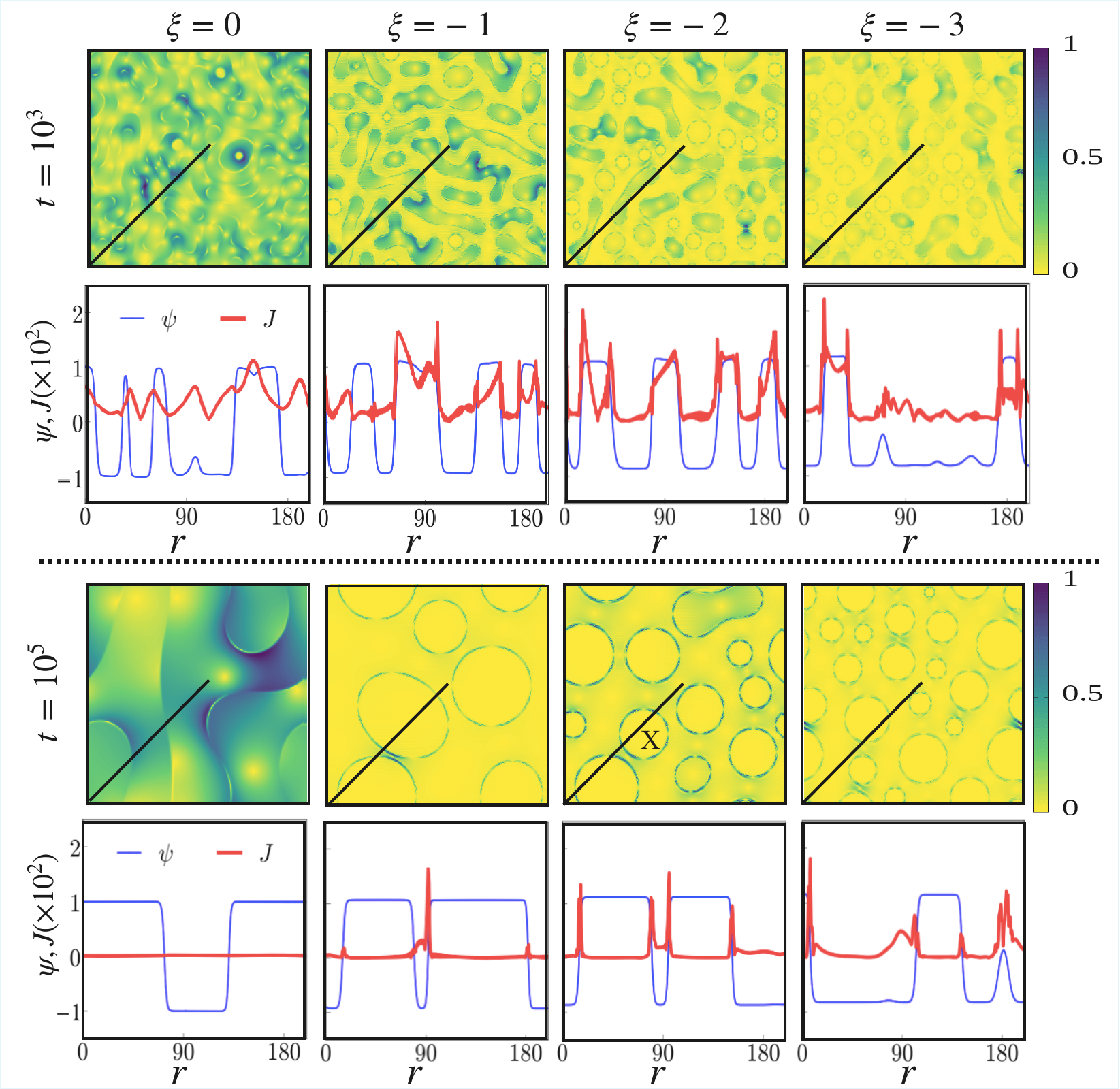} 
\caption{(First row): Snapshots of the current magnitude $J$ for the evolution shown in Fig.~\ref{f1}. The snapshots are taken at $t = 10^3$ for $\lambda = 0$ and $\xi = 0,-1,-2,-3$. The color bar represents the magnitude of $J$, normalized by its maximum value $J_m$ in each snapshot.
(Second row): Profiles of the order parameter $\psi$ and the current magnitude $J$ along the diagonal cross-section shown in the above snapshots. 
(Third and fourth rows): Analogous to first and second rows, but at $t = 10^5$. We will focus on the droplet marked $X$ in the next figure.}
\label{f4}
\end{figure}

To gain a better understanding of the asymptotic growth, let us examine the total current ${\bf J}({\bf r},t)$ that governs transport in AMB+. In Fig.~\ref{f4}, we plot the currents at 2 times ($10^3$ and $10^5$) for the snapshots in Fig.~\ref{f1}. The top row shows the current at $t=10^3$ on a color scale. In the second row, we show the variation of $\psi$ and $J$ (magnitude of ${\bf J}$) along a diagonal cross-section. For nonzero values of $\xi$, the current is very sharply peaked at the domain boundaries due to the dominance of the $\xi$-term in Eq.~(\ref{MBplus}) at the sigmoidal interfaces. This feature becomes more pronounced later, as seen in the snapshots (third row) and profiles (fourth row) for $t=10^5$. The typical interface current at $t=10^5$ is $10^2$-$10^3$ times larger than the bulk current, as we verified by examining the values. 

\begin{figure}
\centering
\includegraphics[width=0.99\linewidth]{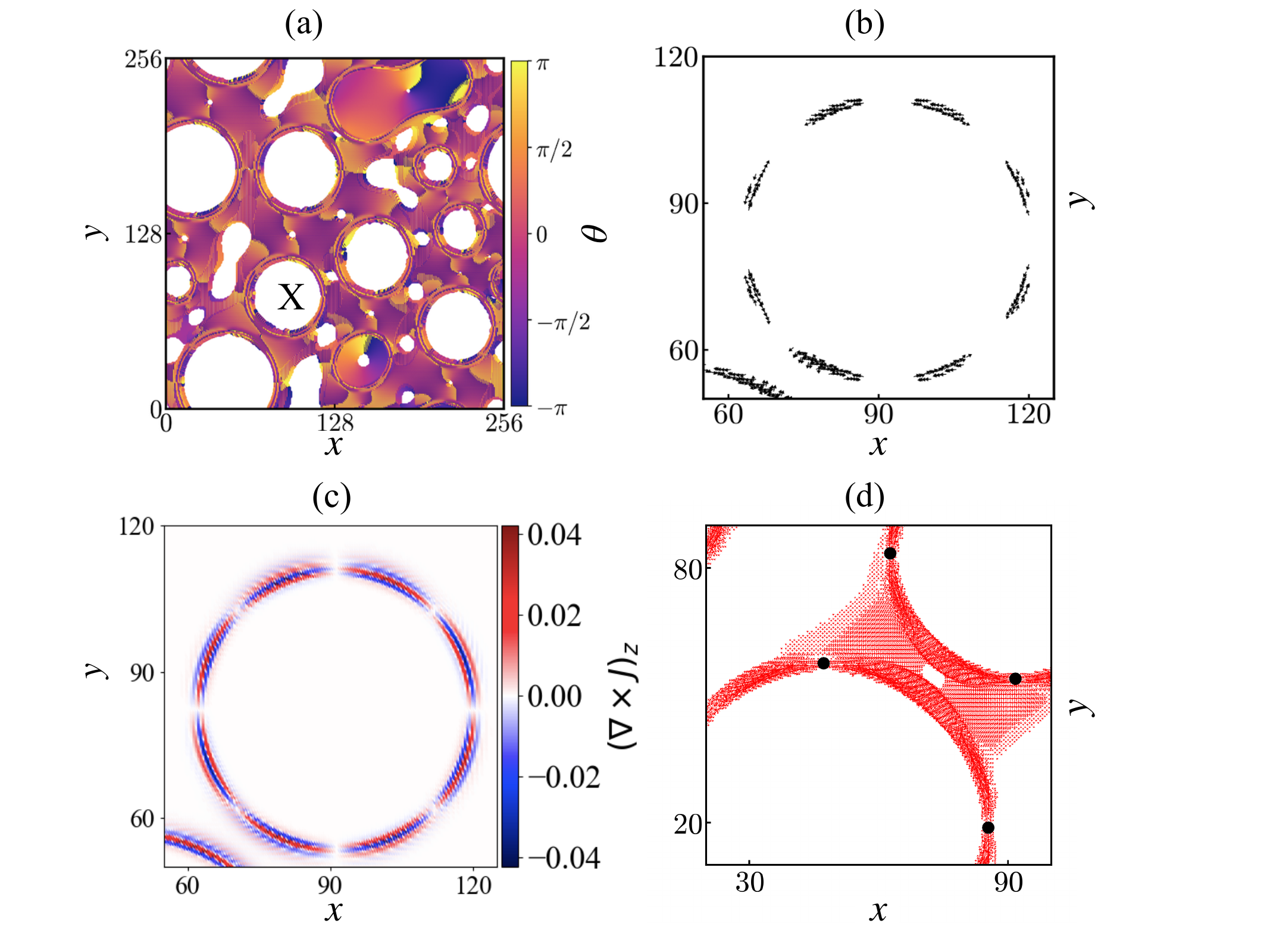} 
\caption{(a) Orientation angle $\theta$ of the current field for $\lambda=0, \xi = -2$ at $t = 10^5$ (see Fig.~\ref{f4}). We show only regions where $J \geq a J_m$, with $a = 0.02$ and $J_m = 0.022$ (maximum current). (b) Arrows indicate the current in regions where $J \geq 0.57 J_m$, showing the direction around the droplet marked $X$ in (a). The length of the arrow is proportional to the current magnitude. (c) Structure of the $z$-component of the curl of the current density $(\nabla \times \mathbf{J})_{\rm z}$ around the droplet marked $X$. The colorbar indicates the magnitude and sign of the curl, with red and blue indicating positive and negative values, respectively. (d) Magnified view of the lower left-hand corner of (b). Arrows denote the current at points where $J\geq 0.05J_m$. Mass moves along a surface loop of the smaller droplet before transferring to a surface loop of the larger droplet. The black dots represent the nodes.}
\label{f5}
\end{figure}

In Fig.~\ref{f5}, we study the direction of the current vector field for $\xi = -2$ at $t = 10^5$. The direction of the vector ($\theta \in [-\pi, \pi]$) is shown using a colormap in Fig.~\ref{f5}(a), where only vectors with $J \geq a J_m$ are shown for clarity ($a = 0.02$, $J_m = 0.022$ is the maximum current magnitude). As expected, the bulk current is negligible, whereas interfacial currents form localized closed loops. A zoomed-in view around a single droplet (marked X) in Fig.~\ref{f5}(b) highlights these circulating flows. To quantify this further, Fig.~\ref{f5}(c) shows the curl of the current, revealing alternating clockwise and anticlockwise current loops along the interface. These loops are separated by nodes where the current is negligible. For droplets of varying sizes, the loops carry currents of different magnitudes. Material is transported from smaller droplets to larger droplets via these unbalanced surface loops, thereby driving forward Ostwald ripening. This is illustrated in Fig.~\ref{f5}(d), where we magnify the lower left-hand corner of Fig.~\ref{f5}(b). The arrows now denote the current at points where $J \geq 0.05 J_m$, i.e., the cut-off is less stringent than in (b) where $J \geq 0.57 J_m$. The total number of vortices is even due to the periodic domain, which enforces zero net vorticity, consistent with Stokes' theorem under periodic boundary conditions.

In MB, ripening is driven by the evaporation of small droplets, diffusive transport through the surrounding region, and condensation on larger droplets. This evaporation-condensation mechanism yields the LS growth law. Thus, tiny currents in bulk domains drive growth. In AMB+, bulk currents are swamped by interfacial currents that drive growth through surface diffusion. In the process of mass transfer, the material has to move along a loop on the surface of a smaller droplet before joining a loop on the surface of a larger droplet. These loops become longer in time, scaling with the droplet size $L$. This tortuous process slows down domain growth as follows. The additional distance covered by the mass is the loop length $\sim L$, renormalizing the effective time to $\tau \sim t/L$, where we have assumed ballistic transport along the loops. For diffusive transport, we expect $L \sim \tau^{1/3}$, yielding the observed growth law $L \sim t^{1/4}$. The same growth law has been derived in the context of MB driven by surface diffusion \cite{pb97,sgs05}.  In Ref.~\cite{pb97}, the role of surface diffusion is amplified by introducing an order-parameter dependent mobility $M(\psi) = 1-\psi^2$ which sets the mobility in the bulk to zero. Such a physical situation arises, e.g., when the bulk phase undergoes gel or glass formation.

Next, we study the dynamical scaling behavior of $C(r,t)$, defined in Eq.~\eqref{crt}. We harden the order parameter field before computing the correlation function by assigning
\begin{eqnarray}
    \psi > 0 & \rightarrow & \psi_1 , \nonumber \\
    \psi < 0 & \rightarrow & \psi_2 .
\end{eqnarray}
This procedure is equivalent to replacing the smooth kinks with step functions. This enables a clear observation of the {\it Porod law} \cite{gp51} that characterizes the scattering from sharp interfaces \cite{op88}. We will discuss this in detail later.

\begin{figure}
\centering
\includegraphics[width=0.9\linewidth]{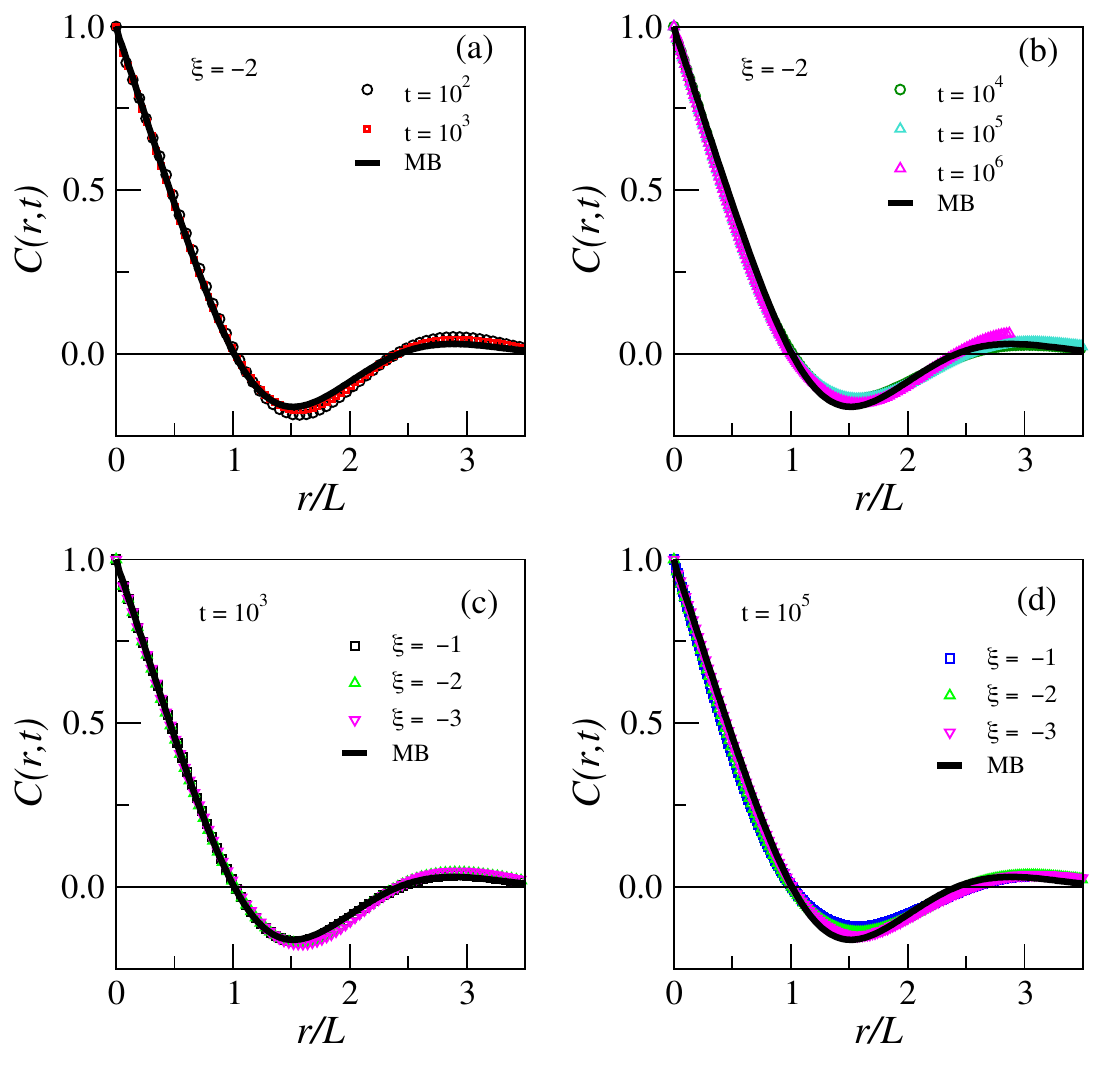} 
\caption{(a) Correlation function $C(r, t)$ vs. $r/L(t)$ for $\lambda= 0, \xi = -2$ at $t = 10^2, 10^3$. The length scale is defined as the distance over which $C(r,t)$ decays to 0. The solid line denotes the scaling function for MB at $t=10^4$. (b) Analogous to (a), but for $t = 10^4, 10^5, 10^6$. (c) Plot of $C(r,t)$ vs. $r/L$ for $\lambda= 0$ and $\xi = -1, -2, -3$ at $t = 10^3$. (d) Analogous to (c), but at $t = 10^5$.}
\label{f6}
\end{figure}

In Fig.~\ref{f6}(a)-(b), we plot $C(r,t)$ vs. $r/L$ for $\xi=-2$. Figs.~\ref{f6}(a) and \ref{f6}(b) show the scaling plot for the early-time regime $t < t_c$ $(t = 10^2, 10^3)$ and the late-time regime $t > t_c$ $(t = 10^4, 10^5, 10^6)$,  respectively.  There is a good data collapse in both regimes. Furthermore, the early-time scaling function is in excellent agreement with the MB function, denoted by a solid line in Fig.~\ref{f6}(a)-(b). However, subtle differences appear in the late-time scaling function of AMB+. For example, the oscillations (dictated by the conservation law $\int d{\bf r}~C({\bf r},t) = 0$) have a smaller amplitude for AMB+. This is a consequence of the crossover in morphology. At first, the morphology of AMB+ is bicontinuous for all $\xi$ as well as MB. Therefore, the scaling is superuniversal, that is, independent of $\xi$. This is confirmed in Fig.~\ref{f6}(c), where we plot the early-time scaling function for $\xi =0$ (MB), $-1, -2, -3$. At late times, the bulk domains in AMB+ saturate to their preferred equilibrium values, which are asymmetric. Due to this asymmetry, the fractions of the two phases are $\phi_1=-\psi_2/(\psi_1-\psi_2)$ and $\phi_2=\psi_1/(\psi_1-\psi_2)$, respectively. Thus, the AMB+ morphology at late times is analogous to the MB morphology with an off-critical composition. For MB, it is well known that the scaling function changes continuously with the off-criticality \cite{op87}, although its analytic form remains an outstanding problem. Thus, we expect the late-time scaling function of AMB+ to continuously vary with $\xi$. This is seen in Fig.~\ref{f6}(d), where we plot the asymptotic scaling function for $\xi = 0,-1,-2,-3$.

In domain growth studies, it is well known that scattering off sharp domain interfaces yields the {\it Porod law} \cite{gp51}. This manifests itself as a linear decay in the correlation function at small $r$ \cite{pw09}:
\begin{equation}
C(r,t) = 1- r/L + \mbox{higher terms} .
\label{porod}
\end{equation}
Equivalently, there is a power-law decay in the tail of the structure factor, which is the Fourier transform of the correlation function:
\begin{equation}
S(k,t) \sim (k L)^{-(d+1)} ,
\end{equation}
where ${\bf k}$ is the momentum and $d$ is the dimensionality \cite{pw09}. The presence of sharp interfaces in AMB+ would naturally yield a Porod law in the present context as well. This is clearly seen in the data for $C(r,t)$ in Figs.~\ref{f6}(a)-(d). The behavior for small $r/L$ is consistent with the linear decay in Eq.~(\ref{porod}).

\begin{figure}
\centering
\includegraphics[width=1.0\linewidth]{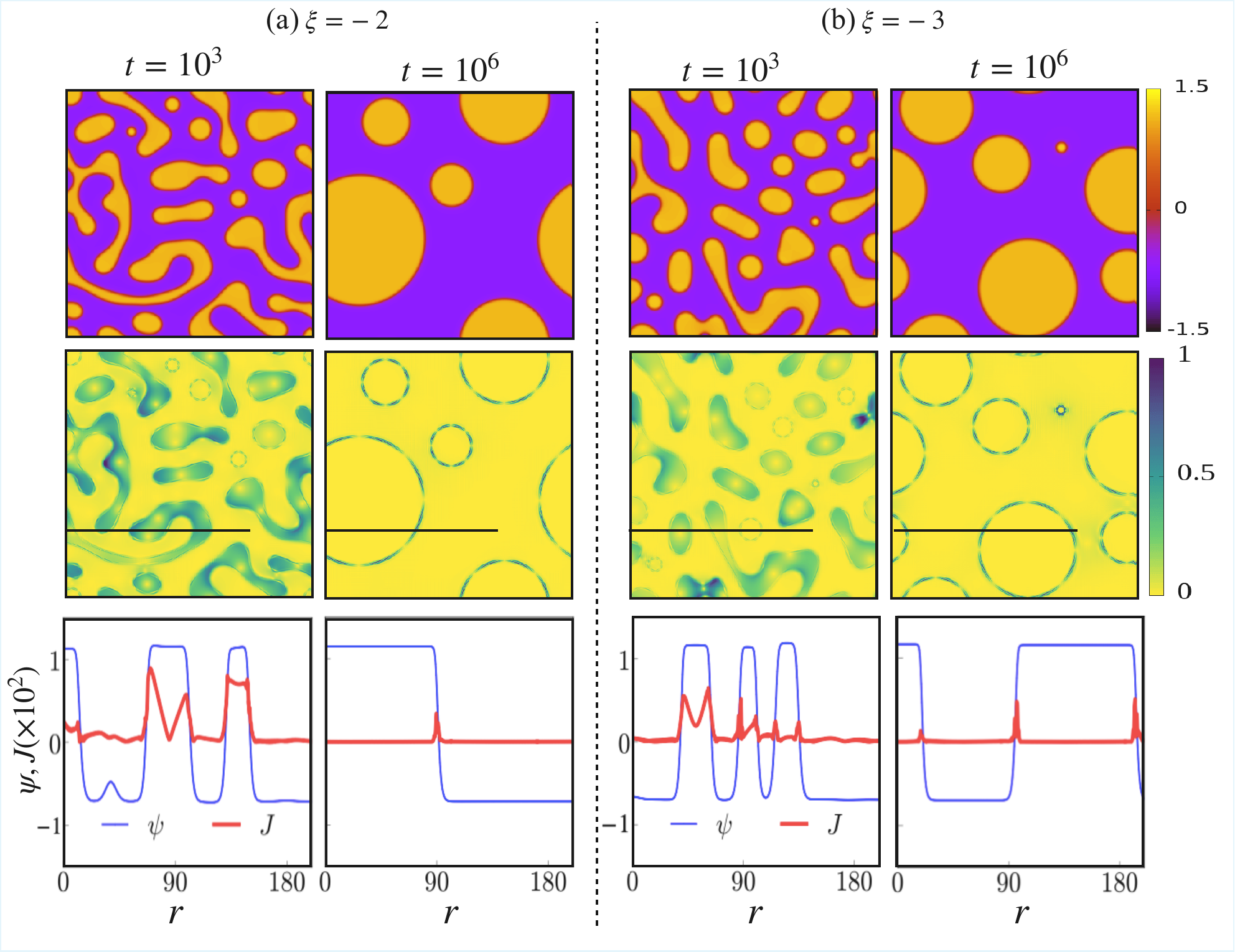} 
\caption{(a) The top row shows snapshots of $\psi$ for $\lambda = 2, \xi = -2$ at $t = 10^3, 10^6 $. The middle row shows $J/J_m$ corresponding to these snapshots. The bottom row shows variation of $\psi, J$ along the black line in the current snapshots. (b) Analogous to (a), but for $\lambda= 2, \xi = -3$.}
\label{f7}
\end{figure}

Before concluding this subsection, let us briefly study MPS for nonzero values of $\lambda, \xi$. We consider the parameter combinations $(\lambda =2, \xi =-2)$ and $(\lambda =2, \xi =-3)$. (These values lie deep in the region labeled forward Ostwald ripening in the phase diagram of \cite{Tjhung2018}.) We have $\alpha = \lambda - \xi/2 > 0$, so we expect $\psi_1 > |\psi_2|$. In Fig.~\ref{f7}, we show evolution snapshots (top row) for these parameters. As before, we see a crossover from the bicontinuous morphology at early times to the droplet morphology at late times. As the area of the $\psi > 0$ phase is smaller than that of the $\psi < 0$ phase, the droplet morphology consists of droplets of $\psi_1$ on a background of $\psi_2$. In the middle frames of Fig.~\ref{f7}, we show a color map of the corresponding current magnitude. As expected, the current peaks sharply at the interfaces. This is validated by the bottom frames of Fig.~\ref{f7}, which show the variation of the current magnitude and the order parameter along a cross section. Upon examination of the numbers, we find that the current at the interfaces is $10^2$-$10^3$ times higher than the bulk current.

Next, we turn to a quantitative characterization of the morphologies in Fig.~\ref{f7}. We have studied the scaling of the correlation function $C(r,t)$. As in the earlier results, we see a crossover in $C(r,t)$ from a superuniversal MB-type form at early times to a parameter-dependent scaling function at late times. For the sake of brevity, we do not show these results here.

\begin{figure}
\centering
\includegraphics[width=0.60\linewidth]{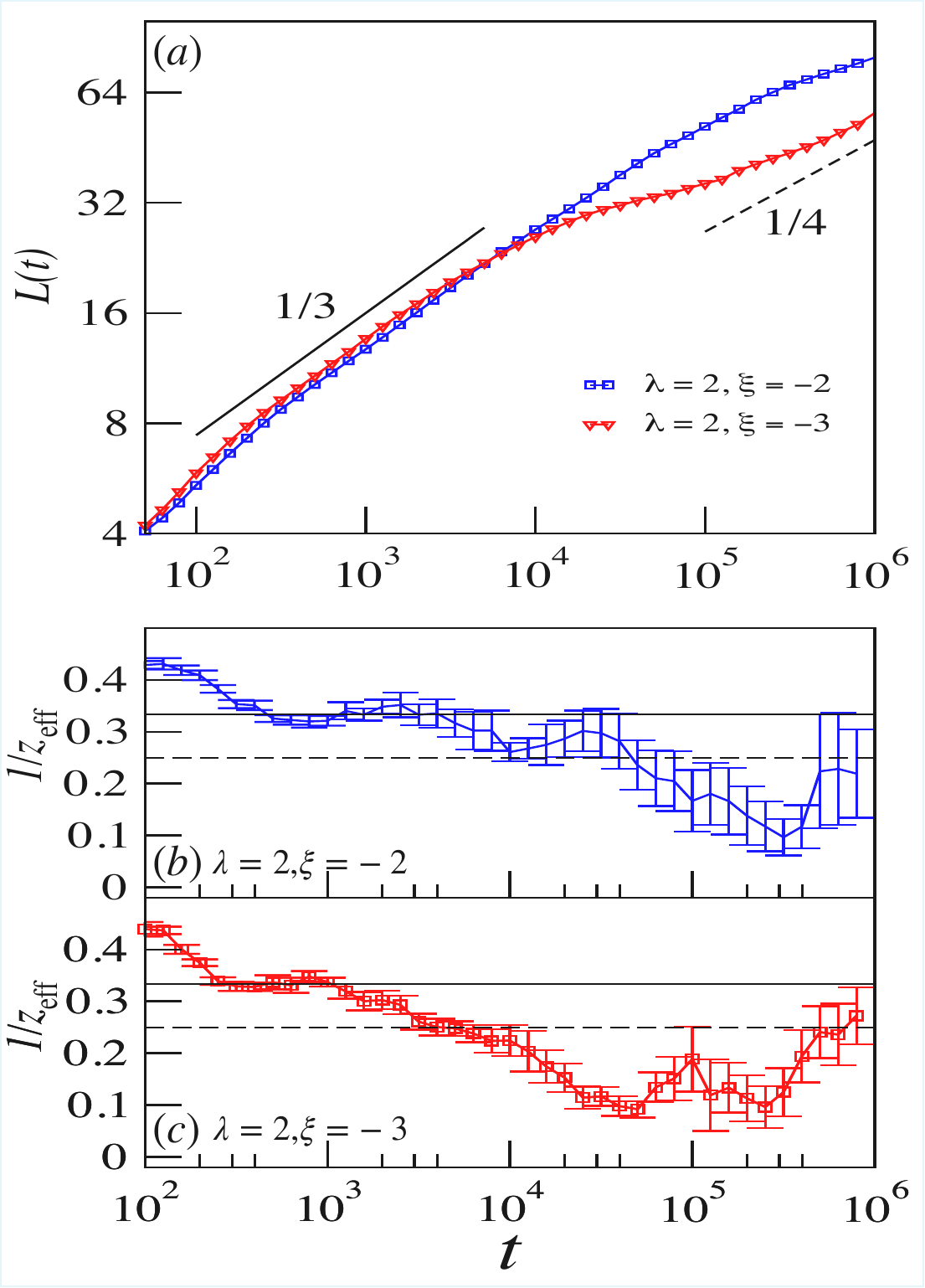} 
\caption{(a) Plot of $L(t)$ vs. $t$ for $\lambda = 2$ and $\xi = -2, -3$ on a log-log scale. The solid line has a slope of $1/3$, while the dashed line has a slope of $1/4$. (b)-(c) Plot of $1/z_{\rm eff}$ vs. $t$ for the data in (a). The horizontal solid and dashed lines denote $1/z_{\rm eff} = 1/3$ and $1/4$, respectively. These data sets are obtained by averaging over 15 independent runs. Hence, the error bars are larger than in Fig.~\ref{f2}.}
\label{f8}
\end{figure}

In Fig.~\ref{f8}(a), we plot $L(t)$ vs. $t$ on a log-log scale for the above parameter values. Because of the dominance of surface diffusion in the late stages, there is a crossover from the LS law to the surface diffusion law at late times. This is confirmed by the plot of the effective exponent ($z_{\rm eff}^{-1}$ vs. $t$) in Figs.~\ref{f8}(b)–(c).

\subsection{Microscale Phase Separation ($\mu$PS)}

\begin{figure}
\centering
\includegraphics[width=0.8\linewidth]{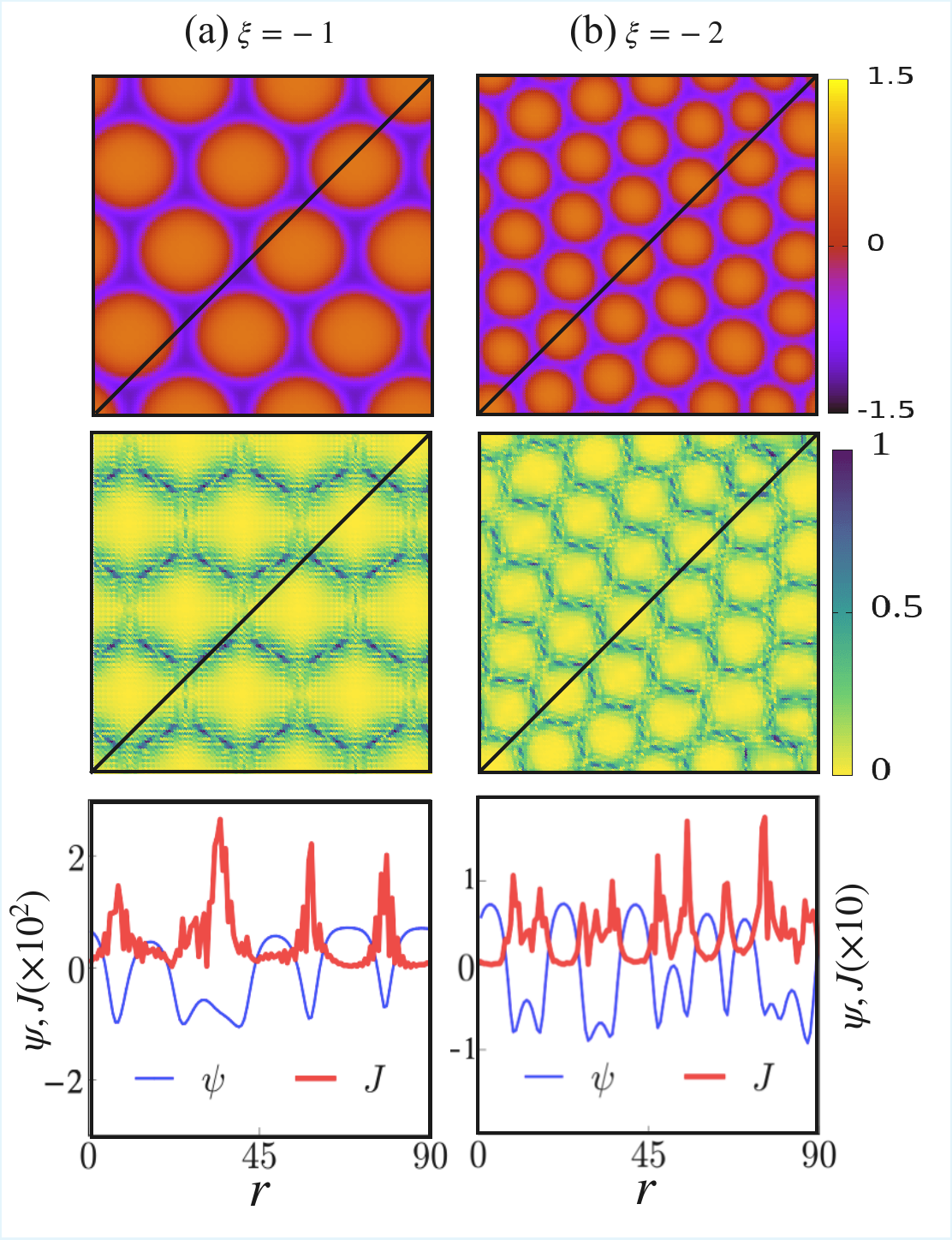} 
\caption{(a) The top frame shows a snapshot of $\psi$ for $\lambda = -4, \xi = -1$ at $t = 10^4$.  The system has a critical composition with $\psi_0 = 0$. The middle frame shows the current magnitude $J/J_m$. The bottom frame shows the variation of $\psi$ and $J$ along the black line in the middle frame. (b) Analogous to (a), but for $\lambda=-4, \xi = -2$. The snapshots are shown for a corner of size $64^2$ to clearly illustrate the spatial arrangement of the droplets.}
\label{f9}
\end{figure}

Next, we present numerical results for $\lambda < 0$ and $\xi < 0$. Our parameter choices lie in the region labeled {\it reverse Ostwald ripening} in the phase diagram of \cite{Tjhung2018}. In Fig.~\ref{f9} (top frames), we show steady-state snapshots at $t = 10^{4}$ for $\lambda = -4$ and $\xi = -1$, $-2$. As in Fig.~\ref{f1}, the color bar indicates the range of $\psi$. As $\alpha=\lambda-\xi/2<0$ in this case, $\psi_1<|\psi_2|$ in Eq.~\eqref{psisat}, so the $\psi<0$ phase is the ``minority phase''. However, the snapshots show static droplets of $\psi_1$ immersed in a background of $\psi_2$. Thus, the droplet morphology in Fig.~\ref{f9} (top frames) is not an off-critical morphology in the sense of MB. Rather, the droplets occupy a larger fraction of the system than the background. 

In Fig.~\ref{f9} (middle frames), we show the magnitude of the current for the above snapshots on a color scale. In the bulk of the droplets, $J$ is almost zero. However, along the surfaces of the droplets, we observe significant currents due to the $\lambda, \xi$ terms in Eq.~(\ref{MBplus}). Again, these form current loops which cancel between adjacent droplets. This balanced configuration ensures that there is no net mass flow between the droplets, thereby maintaining their sizes over time in a steady-state morphology. In Fig.~\ref{f9} (lower frames), we show the order parameter and current along a diagonal section of the snapshots. As argued above, $\psi_1 < |\psi_2|$ as $\alpha < 0$. However, the bulk domains do not saturate to the values $\psi_1$ and $\psi_2$ due to the small droplet sizes. The current magnitude again peaks at the interfaces; however, there is a precise cancellation of clockwise and anticlockwise currents around adjacent droplets.

\begin{figure}
\centering
\includegraphics[width=0.7\linewidth]{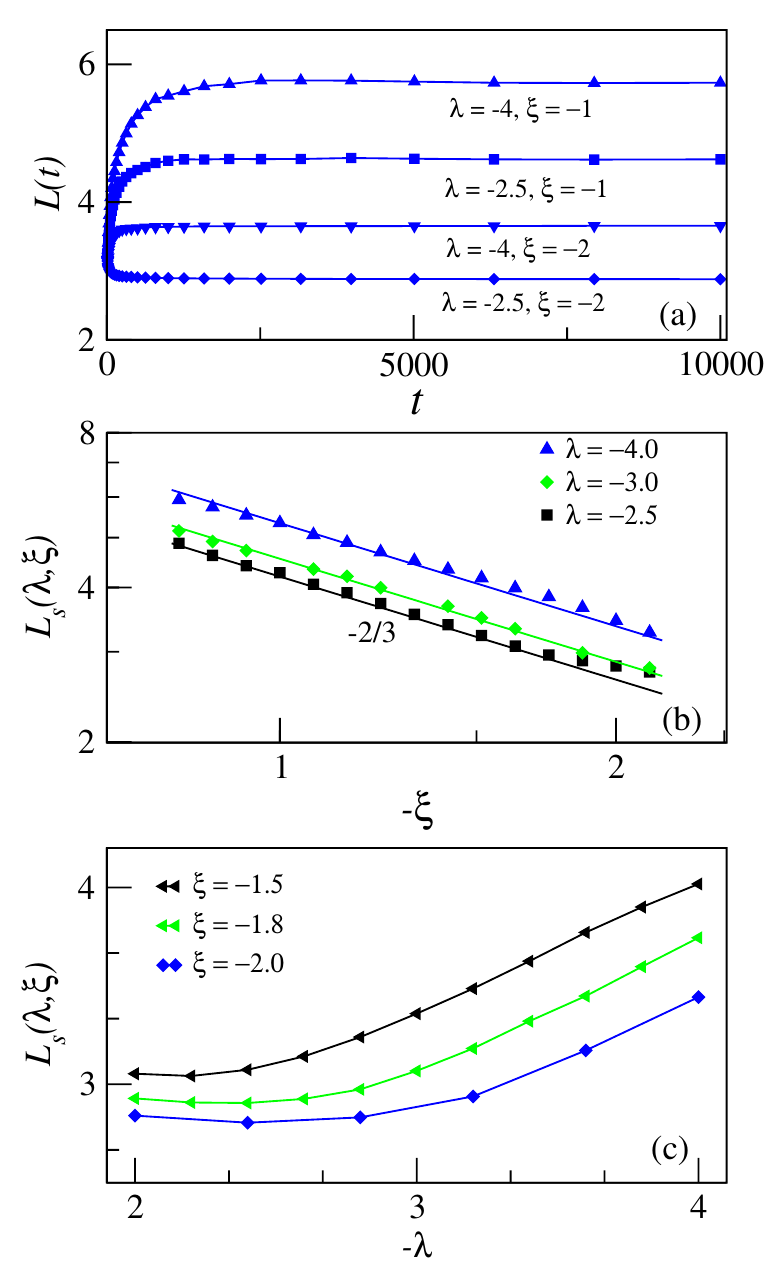} 
\caption{(a) Log-log plot of $L(t)$ vs. $t$ for $\lambda = -2.5, -4$ and $\xi = -1, -2$. (b) Saturation length scale $L_s(\lambda, \xi)$ vs. $-\xi$ for $\lambda = -2.5, -3, -4$. The solid lines represent the best fits to the data with a slope of $-2/3$. (c) Plot of $L_s$ vs. $-\lambda$ for $\xi = -1.5, -1.8, -2$. In all plots, error bars are smaller than the size of the data points.}  
\label{f10}
\end{figure}

Let us quantify the $(\lambda,\xi)$-dependence of the saturation length scale in Fig.~\ref{f9}. We calculate the size of the droplets for several $(\lambda,\xi)$-values. In Fig.~\ref{f10}(a), we plot $L(t)$ vs. $t$ for $\lambda \in \{-2.5, -4\}$ and $\xi \in \{-1, -2\}$. We observe that $L(t)$ grows for early times and saturates to a constant value $L_s$ at late times. The saturated values give the typical droplet size. For a given $\lambda$, the time in which $L(t)$ saturates decreases as $|\xi|$ increases. For a given $\xi$, the saturation time increases for higher values of $|\lambda|$.

Can we make some quantitative statements about the hexagonal morphologies in the upper frames of Fig.~\ref{f9}? First, consider a typical hexagonal unit cell of side $a$, whose vertices are located at the centers of saturated droplets arranged in a regular lattice. Within this cell, we compute the areas occupied by the two phases, $\psi_1$ and $\psi_2$. For a critical composition, the sum of the contributions from both regions must cancel. Therefore, we expect (for $d=2$)
\begin{eqnarray}
     3\pi L_s^2 \psi_1 + \left(\frac{3\sqrt{3}a^2}{2} - 3\pi L_s^2\right)\psi_2 &=&0 .
\end{eqnarray}
Thus, we have
\begin{eqnarray}
a&=&\sqrt{\frac{2\pi(\psi_1-\psi_2)}{\sqrt{3}|\psi_2|}}L_s.
    \label{eq:hexa_side}
\end{eqnarray}
Second, what is the dependence of $L_s(\lambda,\xi)?$ For small values of $\lambda$ or $\xi$, $L_s$ becomes very large. Thus, simulations on any reasonable lattice size encounter finite-size effects before we reach the saturation scale. On the other hand, for large values of $\lambda$ and $\xi$, simulations with the current mesh sizes become numerically unstable. This leaves a limited range of values for $\lambda$ and $\xi$ where we can drive the system to the saturated morphology. In Fig.~\ref{f10}(b), we plot $L_s$ vs. $-\xi$ on a log-log scale for $\lambda =-2.5, -3, -4$. The data appear to be consistent with the power law $L_s \sim (-\xi)^{-2/3}$ in the small $\xi$ range studied. In Fig.~\ref{f10}(c), we plot $L_s$ vs. $-\lambda$ for $\xi= -1.5, -1.8$ and $-2.0$. In this case, $L_s$ shows a more complicated non-monotonic behavior as $L_s \rightarrow \infty$ for $|\lambda| \rightarrow 0$. Simple scaling arguments based on a single droplet do not yield an understanding of the dependence of $L_s(\lambda,\xi)$. We believe that a reasonable interpretation of Figs.~\ref{f10}(b)-(c) requires an understanding of the multi-droplet interactions in Fig.~\ref{f9}.

\begin{figure}
\centering
\includegraphics[width=1\linewidth]{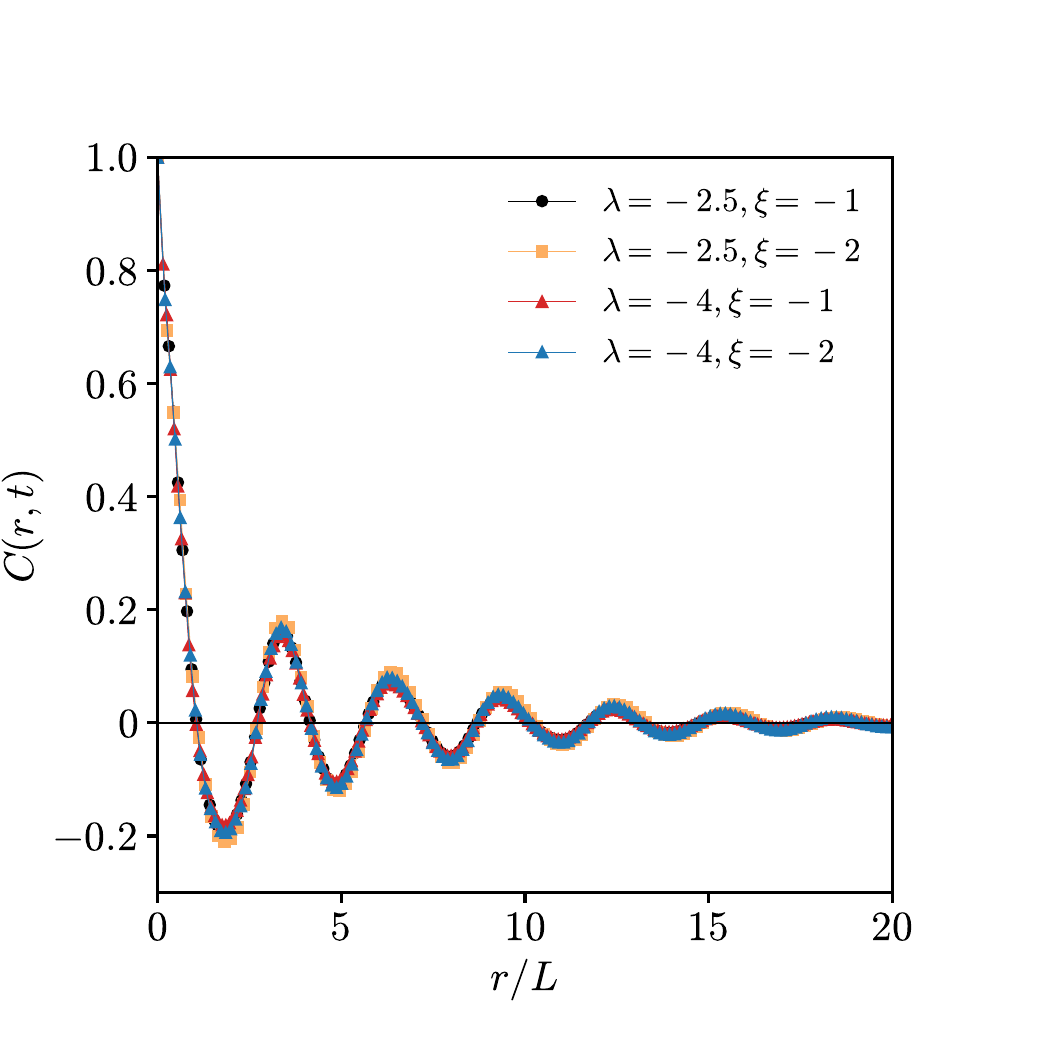} 
\caption{Plot of $C(r,t)$ vs. $r/L$ at $t=10^4$ for $\lambda = -2.5, -4$ and $\xi = -1, -2$.}
\label{f11}
\end{figure}

In Fig.~\ref{f11}, we study the correlation function for the morphologies shown in Fig.~\ref{f9}. We plot $C(r,t)$ vs. $r/L$ at $t=10^4$ (in the saturated regime) for $\lambda = -2.5, -4$ and $\xi = -1, -2$. The scaled correlation function for $\mu$PS is independent of $\lambda$ and $\xi$, and is superuniversal. The corresponding snapshots show a hexagonal morphology for all these $(\lambda, \xi)$-values. The only difference is the magnitude of $L_s$. The scaling function shows marked oscillations, which reflect the crystalline nature of the $\mu$PS morphology.

\section{Summary and Conclusion}
\label{s4}

Let us conclude by summarizing the results of our detailed study of segregation kinetics in {\it Active Model B+} (AMB+), proposed by Tjhung et al.~\cite{Tjhung2018}. AMB+ generalizes the previously introduced {\it Active Model B} (AMB), which consists of Model B (MB) plus an activity term with zero curl and of strength $\lambda$. In AMB+, Tjhung et al.~\cite{Tjhung2018} incorporate an additional active current with non-zero curl and strength $\xi$. We study AMB+ in two cases: (i) only the new rotational term is present $(\lambda=0, \xi \neq 0)$, and (ii) the complete AMB+ with nonzero $\lambda$ and $\xi$. The third case, in which only $\lambda$ is present, was studied earlier in~\cite{sss21}. We study the ordering kinetics of AMB+ after it is ``quenched'' from a homogeneous disordered state to a phase-separated ordered state.

AMB+ has a rich phenomenology and exhibits macro-phase separation (MPS) or micro-phase separation ($\mu$PS) in different regions of $(\lambda,\xi)$ space; see Fig. 4 in Ref.~\cite{Tjhung2018}. In the former case (MPS), the domain scale diverges with a power-law behavior: $L(t)\sim t^{\theta}$. Our numerical studies show that $\theta \simeq 1/4$ in the asymptotic regime. The current peaks sharply at the boundaries in the form of circulating loops, and the dominant growth mechanism is surface diffusion \cite{sgs05,pb97}. The presence of significant currents at the domain interfaces is found in other nonequilibrium systems as well, and is a signature of a system settling into a steady state \cite{NayanaarXiv2025}. In this parameter range, the system evolves by forward Ostwald ripening, with larger droplets growing at the cost of smaller droplets. Because of the distribution of droplet sizes, there is no cancellation of current loops at the droplet boundaries, leading to ongoing macroscopic transport and growth. In addition, evolution morphologies exhibit dynamical scaling, but the asymptotic scaling function depends on $\lambda$ and $\xi$, that is, it is not superuniversal. This is due to the asymmetry between the two phases. Although the overall composition is critical, the morphologies are not bicontinuous. This off-criticality results in the dependence of the scaling function on $\lambda$ and $\xi$. Nevertheless, there are several universal features, e.g., Porod's law due to sharp interfaces.

For the case of $\mu$PS, the emergent morphologies are more subtle. In this case, the kinetics is driven by reverse Ostwald ripening, with smaller droplets growing at the expense of larger droplets. This equalizes the droplet sizes, with a resultant cancellation of circulating currents at the interfaces. In $d=2$, the coarsening system saturates to a morphology in which droplets of size $L_s(\lambda,\xi)$ are located at the vertices of a hexagonal lattice. The lattice spacing has a simple dependence on $L_s$, which is obtained from the conservation law. We believe that a complete understanding of the asymptotic $\mu$PS morphology must account for multi-droplet interactions. Simple arguments based on single-droplet growth are not adequate to understand the dependence of $L_s$ on $\lambda$ and $\xi$. 

For ease of reference, we summarize in Table~\ref{summ} the domain morphologies and coarsening laws for the different models discussed in this paper, highlighting the distinct growth behaviors in AMB, AMB+, and their passive counterpart MB.
\begin{table}
\begin{tabular}{|l|l|l|l|}
    \hline
    Model & Morphology & Growth Exponent & Process \\
    \hline
    MB with MPS & Bicontinuous & $1/3$ & Ostwald ripening \\
    AMB with MPS & Droplet & $1/3 \to 1/4$ & Ostwald ripening \\
    AMB+ with MPS & Droplet & $1/3 \to 1/4$ & Ostwald ripening \\
    AMB+ with $\mu$PS & Droplet & Size saturation & Reverse Ostwald ripening \\
    \hline
\end{tabular}
\caption{Summary of morphologies and growth laws for different models with $\psi_0 = 0$.}
\label{summ}
\end{table}

By now, there have been many theoretical studies that have highlighted the similarities and differences between equilibrium phase separation and nonequilibrium phase separation (as in active matter). The scalar field theories proposed by Cates and coworkers provide a convenient platform for numerical and analytical studies. Unfortunately, our theoretical understanding of even these simple models remains far from complete. In this context, we conducted the studies reported here. We hope that our studies will provoke further interest in this direction. There are also many experiments on this problem, but the results are often contradictory and confusing. Clearly, it would be very valuable to have unambiguous experimental results on the kinetics of active phase separation. The absence of these remains a serious lacuna in understanding the utility of these phenomenological field-theoretical models.

\subsection*{Acknowledgments}

PKY is grateful to the University Grants Commission (UGC), India, for financial support. SM thanks DST-ANRF India for financial support via grants ECR/2017/000659, CRG/2021/006945 and MTR/2021/000438. We are grateful to the referees for their constructive comments and suggestions.

\newpage
\bibliography{ref} 
\newpage

\end{document}